\documentclass{aastex6}

\usepackage{color} 

\fullcollaborationName{}

\begin{document}


\title{Formation of Centaurs' rings through their partial tidal disruption during planetary encounters}


\author{Ryuki Hyodo\altaffilmark{1,2}, S{\'e}bastien Charnoz\altaffilmark{1}, Hidenori Genda\altaffilmark{3} \& Keiji Ohtsuki \altaffilmark{2}}


\altaffiltext{1}{Institut de Physique du Globe, Paris 75005, France}
\altaffiltext{2}{Department of Planetology, Kobe University, Kobe 657-8501, Japan}
\altaffiltext{3}{Earth-Life Science Institute, Tokyo Institute of Technology, Japan}

\begin{abstract}
Centaurs are minor planets orbiting between Jupiter and Neptune that have or had crossing orbits with one or more giant planets. Recent observations and reinterpretation of previous observations have revealed the existence of ring systems around 10199 Chariklo and 2060 Chiron. However, the origin of the ring systems around such a minor planet is still an open question. Here, we propose that the tidal disruption of a differentiated object that experiences a close encounter with a giant planet could naturally form diverse ring-satellite systems around the Centaurs. During the close encounter, the icy mantle of the passing object is preferentially ripped off by the planet's tidal force and the debris is distributed mostly within the Roche limit of the largest remnant body. Assuming the existence of $20-50$wt\% silicate core below the icy mantle, a disk of particles is formed when the objects pass within $0.4-0.8$ of the planet's Roche limit with the relative velocity at infinity $3-6$km s$^{-1}$ and 8h initial spin period of the body. The resultant ring mass is $0.1-10$\% of the central object's mass. Such particle disks are expected to spread radially, and materials spreading beyond the Roche limit would accrete into satellite(s). Our numerical results suggest that ring formation would be a natural outcome of such extreme close encounters and Centaurs can naturally have such ring systems because they cross the orbits of the giant planets.

\end{abstract}




\section{Introduction} \label{sec:intro}
Dynamically unstable Centaurs are minor planets that cross or have crossed orbits with giant planets. The exact definition of the Centaur differs by references. For example, the definition at the Minor Planet Center - MPC/IAU is that the Centaurs are celestial bodies with perihelion beyond the orbit of Jupiter and with semi-major axes smaller than that of Neptune. In contrast, the JPL/NASA defines the Centaurs as objects with semi-major axes between 5.5 au and 30.1 au. Since the discovery of the first Centaur 2060 Chiron \citep{Kow79}, a number of Centaurs have been detected. \cite{Hor04a} estimated the total number of Centaurs with a diameter larger than 1km is approximately 44,000. The sources of the Centaurs are thought to be either the classical Kuiper belt \citep{Lev97}, the scattered disk \citep{Sis07} or the Oort cloud \citep{Bra12}, and their dynamical lifetime is estimated to be $\sim 10^6$ years \citep{Bai09}.\\

Recent occultation observations revealed the existence of narrow ring systems around the Centaurs 10199 Chariklo and 2060 Chiron \citep{Bra14,Ort15,Rup15}. These discoveries prompt us to ask about their dynamical origin as well as their presence only between Jupiter and Neptune. Several models for the formation of ring systems around the Centaurs have been proposed \citep[e.g.][]{Pan16}: (1) Collisional ejection from the parent body's surface, (2) disruption of primordial satellite, and (3) dusty outgassing. For example, \cite{Pan16} discussed ring formation by the lofting of dust particles off Chariklo’s surface into orbit via outflows of sublimating CO, and predicted that rings should be common among large Centaurs but absent among small comets and large Kuiper belt objects. In this work, we propose an alternative dynamical process that can form rings around a Centaur via its extremely close encounters with a giant planet. We perform smoothed particle hydrodynamics (SPH) simulations and show that partial tidal disruption of a differentiated Centaur can form a ring-satellite system around it. In section 2, we briefly discuss the encounter states of Centaurs and their likelihood of encountering giant planets. In section 3, our numerical methods are explained. Numerical results are presented in section 4. Finally, our conclusions and discussions are presented in section 5.

\section{Encounter States}
In this work, we propose that rings around Centaurs could form during a single extremely close encounter with one of the giant planets. The outcome of the close encounter depends on pericenter distance and relative velocity at infinity between the giant planet and the passing object \citep{Hyo15c}. In the following subsections, we will briefly discuss dynamical parameters for the possible encounters of Centaurs with giant planets.

\subsection{Encounter velocity}
\begin{figure}[ht!]
\figurenum{1}
	\plotone{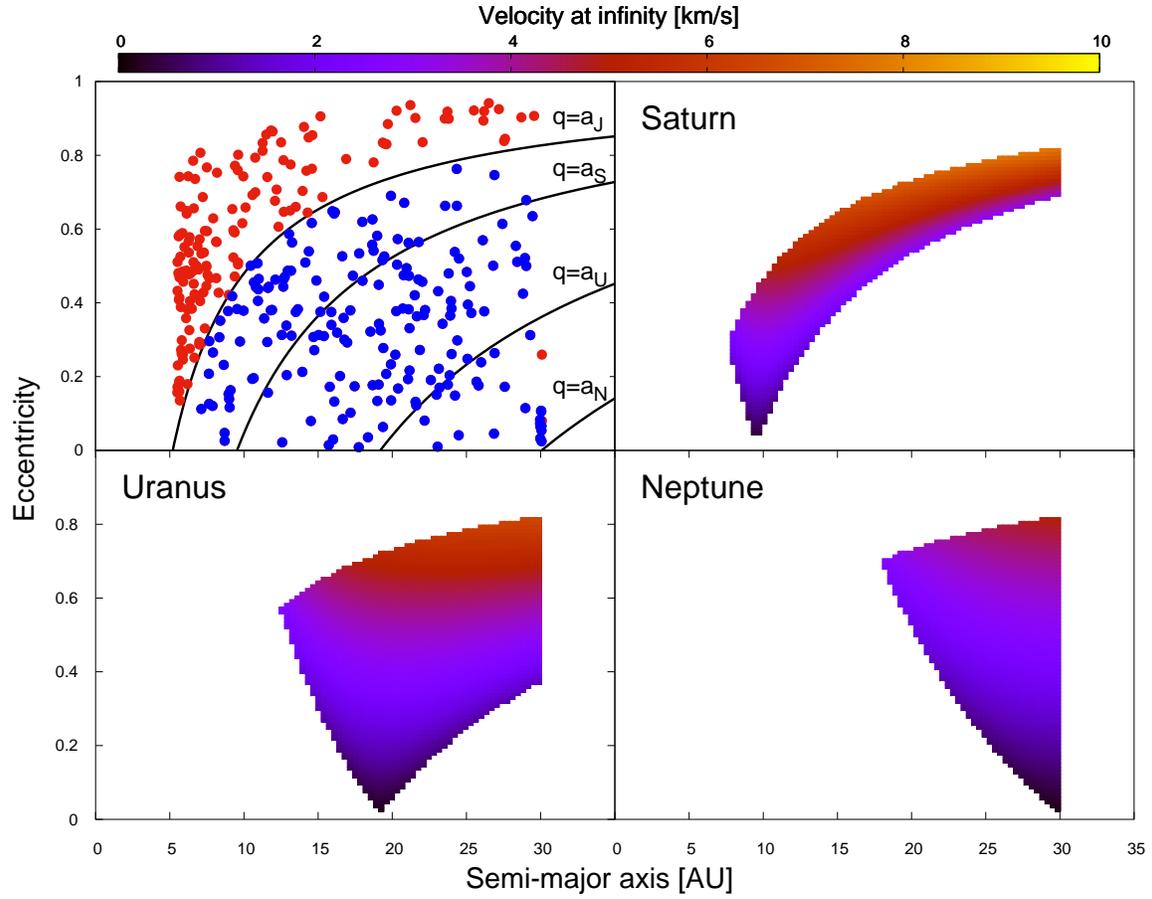}
 \vspace{10mm}
\caption{Semi-major axis eccentricity maps. Dots in the top left panel show the objects that are defined as the Centaurs at JPL/NASA. The blue dots are the Centaurs that are selected using the definition at the Minor Planet Center - MPC/IAU where pericenter distance is larger than the semi-major axis of Jupiter and the semi-major axes are smaller than that of Neptune. Black lines represent the orbits with pericenter distances equal the semi-major axis of Jupiter, Saturn, Uranus and Neptune, respectively. Color contours in the right top and bottom panels are the relative velocities of the Centaur population that follow the MPC/IAU definition at the time of crossing orbit with each giant planet calculated by using equation \ref{rel_vel}.}
\label{figure1}
\end{figure}

Under the gravitational effects of giant planets, the orbits of Centaurs are chaotic and
their orbital elements change during their dynamical lifetime \citep{Tis03,Hor04b}. Thus, the orbital elements of Centaurs can take on a wide range of values on the semi-major axis ($a$)-eccentricity ($e$) plane (Figure \ref{figure1} left top panel). Tidal effects on a passing object at a close encounter depend on the relative velocity at infinity and the distance of the closest approach. Assuming that a giant planet is moving on a circular orbit and a Centaur is orbiting on the same orbital plane with the giant planet, the encounter velocity becomes
\begin{equation}
	v_{\infty}^2 = v_{r}^2 + \left( v_{\rm \theta} - v_{\rm K} \right)^2
\label{rel_vel}
\end{equation}
where $v_r$ and $v_{\theta}$ are the radial and azimuthal components of the velocity of the Centaur at the heliocentric distance $a_0$, and $v_{\rm K}$ is the Keplerian velocity of the giant planet whose semi-major axis is $a_0$, respectively. From the conservation of energy and angular momentum, we have
\begin{equation}
	\frac{1}{2}\left( v_{r}^2 + v_{\rm \theta}^2 \right) - \frac{GM_{\odot}}{a_{0}} =   -\frac{GM_{\odot}}{2a}
\end{equation}
and
\begin{equation}
	\sqrt{GM_{\odot}a\left( 1-e^2 \right)} =  a_{0} \times v_{\rm \theta} 
\end{equation}
where $G$ and $M_{\odot}$ are the gravitational constant and the mass of the Sun, respectively. Then, the velocity components are derived as a function of the orbital elements as
\begin{equation}
	v_{r} = v_{\rm K} \sqrt{ 2 - \frac{a_{0}}{a} - \frac{a(1-e^2)}{a_{0}}}
\end{equation}
and 
\begin{equation}
	v_{\rm \theta} = v_{\rm K} \sqrt{ \frac{a(1-e^2)}{a_{0}} }
\end{equation}

Figure \ref{figure1} shows the encounter velocities of the known Centaurs with either Saturn, Uranus or Neptune. The velocity depends on the orbital elements as well as giant planets, and can take on a variety of values between $v_{\infty} = 0-9$km s$^{-1}$.

\subsection{Probability of encounter with giant planets}
Close encounters with giant planets would occur at two hypothetical epochs; (a) during
the initial encounter with Neptune, it scatters KBOs into the Centaur region, and (b) during the crossing with one of the giant planets after a body is scattered from the Kuiper belt to the Centaur region and becomes a member of the Centaur population. In order to form rings around the Centaur during a close encounter, the Centaur at least needs to pass within the planet's Roche limit. \cite{Ara16} calculated the orbits of Chariklo clones backward and forward in time and show that they can experience multiple extreme close encounters (within $4-5$ planetary radii) with giant planets during their lifetime. \cite{Tis03} calculated the orbits of the known Centaurs and found that $(4\pm2)$\% of the objects impact one of the giant planets. Thus, scaling from their number, the fraction of Centaurs that passes within the Roche limit ($\sim 2$ times of planetary radii)  becomes about $10\%$ or larger. Here we investigate how a close encounter with a giant planet may form a debris disk around a Centaur. 

\section{Numerical Methods and Models}
Using the smoothed particle hydrodynamics (SPH) method, we perform simulations of
close encounters of a differentiated body with a giant planet, where the planet is represented by a point mass. The SPH method is a Lagrangian method \citep{Mon92} in which hydrodynamic equations are solved by considering averaged values of particles through kernel-weighted summation. We ignore the elastic strength, but include the mutual gravity between constituent particles; such an assumption is relevant for bodies larger than several hundred kilometers. We use the Tillotson equation of state \citep{Til62} to calculate the pressure from the internal energy and density. For the artificial viscosity, we use a Von Neumann-Richtmyer-type viscosity with the standard parameter sets ($\alpha=1.0$ and $\beta=2.0$). Our numerical code is the same as that used in \cite{Hyo15c}, which was developed in \cite{Gen12,Gen15a,Gen15b}.\\

According to a recent three-dimensional thermal evolution model, trans-Neptunian objects and Centaurs could be strongly differentiated and forming stratified internal structures due to the heating by short-lived radioactive isotopes whose amount depends on the timing of their formation \citep{Gui11}. If the Centaurus formed within 1 Myr, the body with a radius larger than 10 km should be differentiated \citep{Gui11} In this work, we assume that the object is differentiated with either 50wt\% or 20wt\% of a silicate core covered by a water ice mantle with a total mass of $10^{20}$kg. The object consists of 100,000 SPH particles. Initial positions and velocities of all the particles that follow a hyperbolic orbit around Saturn are given analytically with initial distance to the planet being set to $3.0 \times 10^5$ km, which is about twice as long as the Roche limit of water ice material. This starting point is large enough to neglect the tidal effect of the planet \citep{Hyo14}. The hyperbolic orbit is entirely determined by the pericenter distance $q$ and the velocity at infinity $v_{\infty}$. In this work, we explore different values of pericenter distances within the planet's Roche limit, and the values of the velocity at infinity are chosen from the range $v_{\infty} = 3.0$km s$^{-1}$ or $6.0$km s$^{-1}$, which are expected for the known Centaurs (see Figure \ref{figure1}). We also investigate the dependence on the initial spin state of the passing objects. The spin-axis of the body is assumed to be perpendicular to the orbital plane with either prograde or retrograde spin to the direction of the hyperbolic orbit with the spin period $P=8$h. Such a rotation period is common in the trans-Neptunian belt \citep{Thi14}. The critical distance for tidal disruption can be scaled by the planet's Roche limit \citep{Sri92}. Thus, the dependence on the pericenter distance here is shown by scaling with the planet's Roche limit $a_{\rm R}$ for mantle material, so that our results can also be applicable to other giant planets.\\

As shown in the next section, part of the material scattered around the object after the close encounter becomes gravitationally bound around the central body (the largest remnant produced after the encounter), and these bound particles have highly eccentric orbits immediately after the encounter. However, collisions between the particles rapidly damp the eccentricities and inclinations and form a disk of particles on nearly circular orbits while conserving the angular momentum. Thus, we calculate the equivalent circular orbital radius $a_{\rm eq}=a(1-e^2)$, where $a$ and $e$ are the post-encounter semi-major axis and eccentricity of a disk particle, respectively. Then, the total mass of the particles that is gravitationally captured by the central body and satisfies $a_{\rm eq}>R_{\rm body}$ ($R_{\rm body}$ is the radius of the body) is regarded as the ring mass. Our numerical simulations are stopped when the disk mass becomes almost constant.\\

\section{Ring-satellite formation during extreme close encounters}
\begin{figure}[ht!]
\figurenum{2}
	\plotone{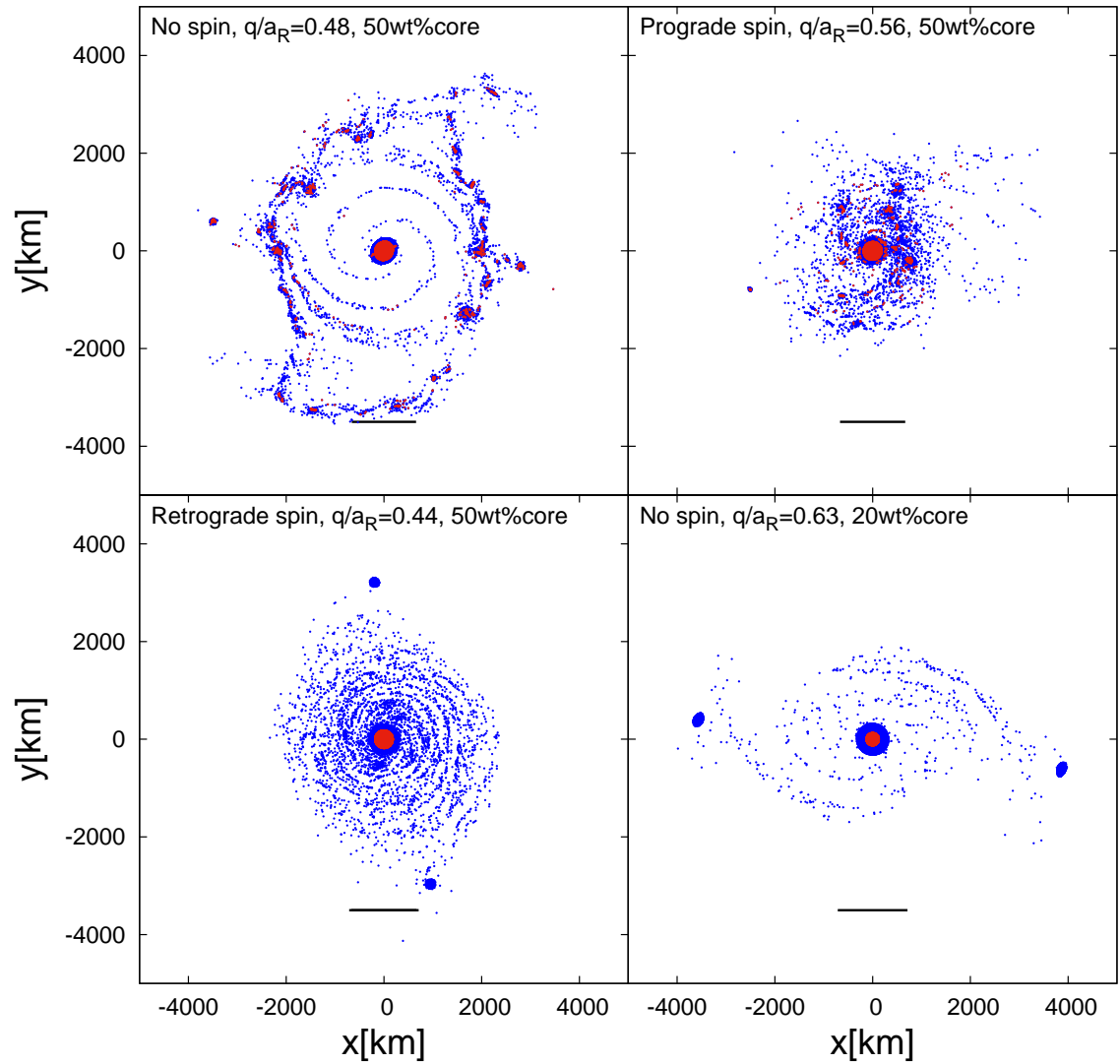}
 \vspace{15mm}
\caption{Snapshots of simulations for the different initial states and models with $v_{\infty}=3$km s$^{-1}$. Left top panel shows the case of initial no-spin, $q/a_{\rm R}=0.48$ with a 50wt\% silicate core. Right top panel shows the case of initial prograde spin, $ q/a_{\rm R}=0.56$ with a 50wt\% silicate core. Left bottom panel shows the case of initial retrograde spin, $ q/a_{\rm R}=0.44$ with a 50wt\% silicate core. Right bottom panel shows the case of initial no-spin, $ q/a_{\rm R}=0.63$ with a 20wt\% silicate core. Blue and red colors represent icy and silicate components, respectively. The horizontal black lines are the Roche diameter of the central objects for the ice material.}
\label{figure2}
\end{figure}

During an extreme close encounter with a giant planet, the tidal force significantly spins up in a prograde direction and deforms the passing object. When an object passes too deep inside the potential field of the giant planet, strong tides homogeneously stretch and destroy not only the mantle, but also the core that splits into several small clumps of similar size without forming a ring around a central remnant \citep{Hyo15c}. However, when objects experience weaker tides at slightly larger pericenter distance, the core is not fully destroyed and only the icy mantle is preferentially stripped off from the surface and distributed around the largest remnant, resulting in the formation of a ring that is a mixture of ice and silicate. The outcomes of this destruction are diverse depending on the initial state (internal structure, spin) of the incoming body and the geometry of the encounter (Figure \ref{figure2}). We performed additional SPH simulations for the case of undifferentiated bodies. We found that ring formation is difficult for the undifferentiated bodies, since the Roche limit of the largest remnant becomes too small to capture debris during tidal disruption at the close encounter.\\

\begin{figure}[ht!]
\figurenum{3}
	\plotone{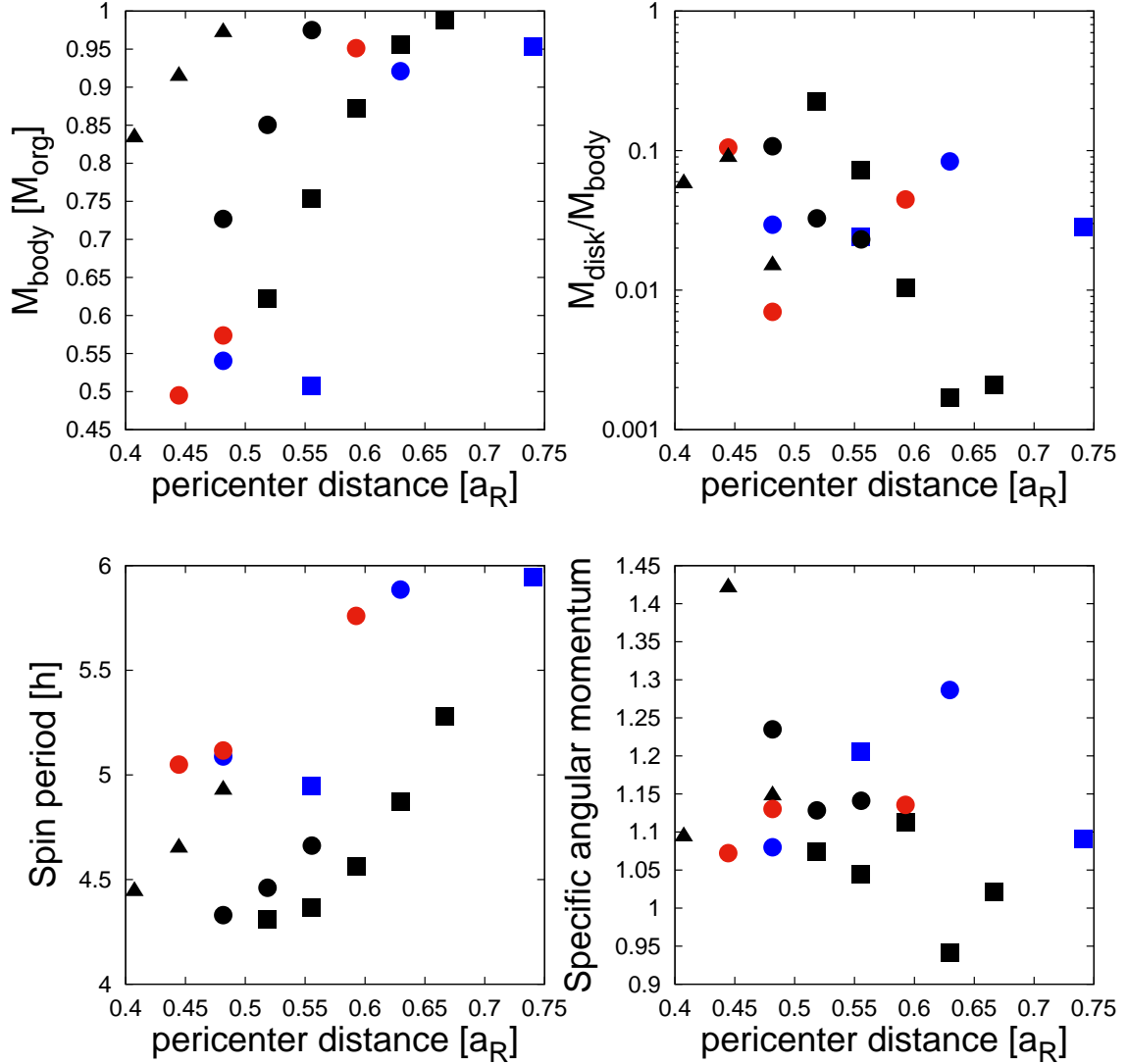}
 \vspace{15mm}
\caption{Properties of the largest remnant body and the particle disk formed around it by tidal disruption during a close encounter of a differentiated body with a giant planet. Horizontal axis is the pericenter distance of the initial orbit of the passing body scaled by the Roche limit of the planet for icy material (135,000km in the case of Saturn). Top left and right panels show the mass of the central largest remnant $M_{\rm body}$ scaled by the initial mass of the passing body $M_{\rm org}$, and the disk mass $M_{\rm disk}$ scaled by that of the central largest remnant, respectively. Bottom left panel shows the post-encounter spin period of the central largest remnant, and the bottom right panel shows the specific angular momentum of the particle disk scaled by $M_{\rm disk}\sqrt{GM_{\rm body}a_{\rm R,lr}}$, where $a_{\rm R,lr}$ is the Roche limit of the central object. Different symbols represent different initial spin states; circles, triangles and squares represent cases of no spin, retrograde and prograde spins, respectively. Black points are those of $v_{\infty}=3$km s$^{-1}$ with a 50wt\% core. Red and blue points are those of a 20wt\% core with $v_{\infty}=6$km s$^{-1}$ and $v_{\infty}=3$km s$^{-1}$, respectively. Note that the radii of giant planets scaled by their Roche limits for the mantle material are 0.43, 0.36 and 0.33 for Saturn, Uranus and Neptune, respectively.}
\label{figure3}
\end{figure}

Within our limited calculations with $v_{\infty}=3$ or $6$km s$^{-1}$, 20wt\% or 50wt\% silicate core and initial spin period $P=8$h in either prograde or retrograde direction, when the body passes at a distance between $0.4-0.8 a_{\rm R}$, the outcome is ringed objects. The resultant ring's mass is $0.1-10$\% of the mass of the central object and mostly distributed within the Roche limit (Figure \ref{figure3}). Within the range of these parameters, when the tides are strong enough (inner edge of the parameter range at each combination of $v_{\infty}$, spin period and internal structure), even core material is incorporated into the ring (Figure \ref{figure2} top panels) in addition to the icy mantle. In contrast, when the tides are slightly weaker, only the icy mantle is destroyed and pure icy rings or disks are formed (Figure \ref{figure2} bottom panels). Furthermore, not only does this form debris rings, but, in some cases, part of the progenitor's mantle is also directly split into small icy moons (Figure \ref{figure2} bottom panels). Note that, these moons usually have highly eccentric orbits around the central object, and thus investigation on their longer-term evolution is required in order to understand the successive dynamical evolution of their ring-satellite systems.\\

As the pericenter distance becomes larger, the outcome is less destructive, generally forming less massive disks (see black circle or square points in Figure \ref{figure3} right top panel). However, when moons are directly formed by splitting of the mantle, larger total mass (disk mass + mass of the moons) orbits around the central object (see Figure \ref{figure2} bottom panels and the corresponding black triangle $q/a_{\rm R}=0.44$ or blue circle $q/a_{\rm R}=0.63$ in Figure \ref{figure3} right top panel). Such direct moon formation might be the typical outcome in the case of a small core and large pericenter distance. The outcome also depends on the internal structures, initial spin states and velocity at infinity. For a smaller fraction of the silicate core, faster prograde spin or smaller value of velocity at infinity, the encounter becomes more destructive, and thus ring formation can occur at larger pericenter distances. During the close encounter, the object is spun up in the prograde direction and as the pericenter distance becomes larger, the object is less spun-up (Figure \ref{figure3} left bottom panel). The critical spin period against its self-gravity is $P_{\rm crit}=2\pi / \Omega_{\rm crit}$ where $\Omega_{\rm crit}=\sqrt{4\pi \rho G/3}$, and $P_{\rm crit} \sim 3.8$h and $2.2$h for water ice ($\rho=900$kg m$^{-3}$) and silicate ($\rho=2700$kg m$^{-3}$), respectively. The resultant spin periods of the largest bodies in our simulations are about $4-6$h (Figure \ref{figure3} bottom left panel). Interestingly, these final spin periods are rather similar to those of Chariklo and Chiron, which are $P\sim7$h and $5.5$h, respectively.\\

After their formation, the rings are expected to spread and accrete into moon(s) outside the Roche limit of the hosting body \citep{Cha10,Cri12,Hyo15a}. Then, collisions between formed moons just outside the Roche limit would result in the formation of a narrow ring and shepherd moons as in the case of the formation of Saturn's F ring and its shepherds \citep{Hyo15b}. Alternatively, moons formed by directly splitting from the progenitor's mantle during the encounters might shape the debris rings in a longer-term evolution. Such successive dynamical evolution may explain the observed narrow ring structures around Chariklo and Chiron. However, more detailed direct numerical simulations are required to fully understand the stochastic nature of the ring-satellite system evolution \citep[see also][]{Hyo15a}.\\

\section{Discussions \& Conclusions}
The Centaurs are transient objects that cross the orbits of one or several giant planets. During their short dynamical lifetime ($\sim 10^6$ years), they frequently undergo close encounters with giant planets. Motivated by recent discoveries of the ringed large Centaurs, 10199 Chariklo and 2060 Chiron, we performed SPH simulations in order to investigate possible outcomes of tidal disruption of a differentiated large Centaur during an extremely close encounter with a giant planet.\\

We found that the outcomes of the close encounters depend on the initial spin state and the internal structure of the passing object. When the passing body has a 20wt\% or 50wt\% dense core, an initial spin period $P=8$h in either prograde or retrograde direction and its initial orbit has a pericenter distance of $0.4-0.8a_{\rm R}$ with $v_{\infty}=3$ or $6$km s$^{-1}$, our SPH simulations show that the close encounter naturally results in the formation of debris rings around the largest remnant. In such cases, the icy mantle is preferentially stripped off during the close encounter and mostly distributed within the Roche limit of the largest remnant without significant disruption of the silicate core. Thus, the resultant ring composition becomes either pure water ice or the mixture of water ice and a small amount of silicate. In some cases, not only the formation of the rings but also small moons are formed directly by splitting from the mantle or the core. In order to form rings, the tidal force should not be too strong or too weak. When it is too strong, the objects are homogeneously stretched/destroyed and form similar sized smaller clumps \citep{Hyo15c}. In contrast, when the tidal force is too weak, the passing body remains intact. The outcome also depends on the internal structure of the passing body. We find that the existence of a dense core helps the formation of the rings. Without the core, the outcome tends to be splitting into several clumps. Thus, if our scenario is correct, Chariklo and Chiron should be differentiated. However, more detailed investigations are necessary to fully understand the range of initial orbital parameters as well as the internal structure of the passing Centaurs that can form ringed remnant objects after the close encounters. We will leave these questions to our future work. As discussed in Section 2.2, simple analytical estimation provides us about 10\% of all Centaurs can boast ring-satellite systems. However, current observation is not enough to have a good statistic. So far, we know Chariklo and Chiron are only ringed Centaurs. However, future occultations may be able to detect more rings or orbiting features.\\

In our present calculations, we found that the mass of the formed rings is $0.1-10$\% of the mass of the largest remnant body and the disk materials are mostly distributed within the Roche limit of the body. Such rings spread, forming satellite(s) outside the Roche limit \citep{Ida97,Kok00,Cha10,Cri12,Hyo15a}. Furthermore, collisional disruption between formed moons just outside the Roche limit would create a narrow ring and shepherd satellites \citep{Hyo15b}. Therefore, the currently observed/suggested narrow ring systems around Chariklo or Chiron might be the outcomes of such dynamical evolution after the ring formation by tidal disruption. Alternatively, moons that would have formed directly by splitting from the icy mantle or cores would have shaped the remnant rings. Therefore, in addition to rings as revealed by recent observations for Chariklo and Chiron, our results suggest that a significant fraction of Centaurs should have small moon(s) awaiting discovery. Direct simulations of the longer-term evolution are required to clarify the evolution of the ring-satellite system.\\

\acknowledgments
This work was supported by JSPS Grants-in-Aid for JSPS Fellows (15J02110) and for Scientific Research B (22340125 and 15H03716). We acknowledge the financial support of the UnivEarthS Labex program at Sorbonne Paris Cit{\'e} (ANR-10-LABX-0023 and ANR-11-IDEX-0005-02). This work was also supported by Universit{\'e} Paris Diderot and by a Campus Spatial grant. S{\'e}bastien Charnoz thanks the IUF (Institut Universitaire de France) for financial support.


\begin{thebibliography}{}
\bibitem[Araujo et al.(2016)]{Ara16} Araujo, R.A.N., Sfair, R. \& Winter, O.C., 2016, \apj, 824, 80
\bibitem[Bailey \& Malhotra(2009)]{Bai09} Bailey, B. L., \& Malhotra, R. 2009, Icarus, 203, 155
\bibitem[Brasser et al.(2012)]{Bra12} Brasser, R., Schwamb, M. E., Lykawka, P. S., et al. 2012, MNRAS, 420, 3396
\bibitem[Braga-Ribas et al.(2014)]{Bra14} Braga-Ribas, F., Sicardy, B., Ortiz, J. L., et al. 2014, Nature, 508, 72
\bibitem[Charnoz et al.(2010)]{Cha10} Charnoz, S., Salmon, J., \& Crida, A. 2010, Nature, 465, 752
\bibitem[Crida \& Charnoz(2012)]{Cri12} Crida, A., \& Charnoz, S. 2012, Science, 338, 1196 
\bibitem[Di Sisto \& Brunini(2007)]{Sis07} Di Sisto, R. P., \& Brunini, A. 2007, Icarus, 190, 224
\bibitem[Genda et al.(2012)]{Gen12} Genda, H., Kokubo, E., \& Ida, S. 2012, ApJ, 744, 137
\bibitem[Genda et al.(2015a)]{Gen15a} Genda, H., Fujita, T., Kobayashi, H., et al. 2015, Icarus, 262, 58-66
\bibitem[Genda et al.(2015b)]{Gen15b} Genda, H., Kobayashi, H., \& Kokubo, E. 2015, ApJ, 810, 136
\bibitem[Guilbert-Lepoutre et al.(2011)]{Gui11} Guilbert-Lepoutre, A., Lasue, J., Federico, C., Coradini, A., Orosei, R., Rosenberg, E.D., 2011, A\&A, 529, 71
\bibitem[Hyodo \& Ohtsuki(2014)]{Hyo14}  Hyodo, R., \& Ohtsuki, K., 2014, \apj,  787, 56
\bibitem[Hyodo et al.(2015)]{Hyo15a} Hyodo, R., Ohtsuki, K.\& Takeda, T. 2015, \apj, 799, 40
\bibitem[Hyodo \& Ohtsuki(2015)]{Hyo15b} Hyodo, R., \& Ohtsuki, K. 2015,  Nature Geo., 8, 686-689
\bibitem[Hyodo et al.(2016b)]{Hyo15c} Hyodo, R., Charnoz, S., Ohtsuki, K., \& Genda, H. 2016b Icarus, submitted
\bibitem[Horner et al.(2004a)]{Hor04a} Horner, J., Evans, N.W., \& Bailey, M.E., 2004a, MNRAS, 354, 798-810
\bibitem[Horner et al.(2004b)]{Hor04b} Horner, J., Evans, N.W., \& Bailey, M.E., 2004b, MNRAS, 355, 321-329
\bibitem[Ida et al.(1997)]{Ida97} Ida, S., Canup, R. M., \& Stewart, G.R. 1997, Nature, 389, 353
\bibitem[Kokubo et al(2000)]{Kok00} Kokubo, E., Ida, S., \& Makino, J. 2000, Icarus, 148, 419
\bibitem[Kowal et al.(1979)]{Kow79} Kowal C. T., Liller W., Marsden B. G., 1979, Proc. IAU Symp. 81, Dynamics of the Solar System. Reidel, Dordrecht, p. 245
\bibitem[Levison \& Duncan(1997)]{Lev97} Levison, H. F., \& Duncan, M. J. 1997, Icarus, 127, 13
\bibitem[Monaghan(1992)]{Mon92} Monaghan, J.J., 1992 Annu. Rev. Astron. Astrophys., vol. 30 (A93-25826 09-90) 30, 543-574
\bibitem[Ortiz et al.(2015)]{Ort15} Ortiz, J. L., Duffard, R., Pinilla-Alonso, N., Alvarez-Candal, A., Santos-Sanz, P., Morales, N., Fernndez-Valenzuela, E., Licandro, J.,Campo Bagatin, A. \& Thirouin, A., 2015, Astron. \& Astrophys., 576, 18
\bibitem[Pan \& Wu(2016)]{Pan16} Pan, M., P. \& Wu, Y., 2016  Astron. \& Astrophys., 821, 1
\bibitem[Ruprecht et al.(2015)]{Rup15} Ruprecht, J. D., Bosh, A. S., Person, M. J., Bianco, F. B., Fulton, B. J., Gulbis, A. A. S., Bus, S. J., \& Zangari, A. M. 2015, Icarus, 252, 271
\bibitem[Sridhar \& Tremaine(1992)]{Sri92} Sridhar, S., \& Tremaine, S. 1992. Icarus, 95, 86-99
\bibitem[Tillotson(1962)]{Til62} Tillotson, J.H., 1962 General atomic rept. GA-3216, 1-142
\bibitem[Thirouin et al.(2014)]{Thi14} Thirouin, A., K.S. Noll, J.L. Ortiz, \& N. Morales 2014, Astron. \& Astrophys., 569, A3.1-20
\bibitem[Tiscareno \& Malhotra(2003)]{Tis03} Tiscareno, M. S., \& Malhotra, R. 2003, \aj, 126, 3122
\end{thebibliography}
\end{document}